\documentstyle[preprint,aps]{revtex}
\input{epsf}
\def\aa{\hat{a}}
\begin{document}

\title{Interfacial Velocity Corrections due to Multiplicative Noise}

\author{Leonid Pechenik and Herbert Levine}
\address{Dept. of Physics \\
University of California, San Diego \\ La Jolla, CA  92093-0319}
\maketitle
\address{
Department~of~Physics, University~of~California,~San~Diego, 9500~Gilman~Drive,
La Jolla, CA~92093-0354, USA
}

\maketitle

\begin{abstract}
The problem of velocity selection for reaction fronts has been intensively
investigated, leading to the successful marginal stability approach for
propagation into an unstable state. Because the front velocity is controlled
by the leading edge which perforce has low density, it is interesting to
study the role that finite particle number fluctuations have on this
picture. Here, we use the well-known mapping of discrete Markov processes
to stochastic differential equations and focus on the front velocity in
the simple $A+A \stackrel{\leftarrow}{\rightarrow} A$ system. Our results
are consistent with a recent (heuristic) proposal that 
$v_{MS} - v \sim {1\over \ln^2 {N}}$.
\end{abstract}


\section{Introduction}

There has been a great deal of interest in the problem of reaction
front propagation in non-equilibrium systems. This issue arises in
systems ranging from flames~\cite{kpp} to bacterial colonies~\cite{bacteria}, from solidification
patterns~\cite{review} to genetics~\cite{genetics}. Most of the 
theoretical work in this
area involves solving deterministic reaction-diffusion equations. Here,
we focus on effects that occur when one goes beyond this mean-field
treatment and considers the effects of fluctuations.

By now, it is clear
that there are several possible mechanisms whereby the velocity of
a deterministic reaction-diffusion front can be selected. For cases where
we propagate into a linearly unstable state, the marginal stability
criterion~\cite{marginal} suggests that the fastest stable front is the one that
is observed for all physical initial conditions. For propagation into
a metastable state, there is a unique front solution consistent with
the boundary conditions and hence there is no selection to be done. In
between, there is the case of a nonlinearly unstable state in which the
exponentially localized front is chosen. These principles have been
verified in many examples and in some cases can be rigorously 
derived~\cite{Aronson}.

However, it is understood that deterministic 
equations are often only approximations to the actual non-equilibrium
dynamics. This is particularly clear in the case of chemical reaction
systems where the true dynamics is a continuous time Markov process which
gives rise to a reaction-diffusion equation only in the limit of an infinite
number of particles per site~\cite{Kampen}. More generally, having a finite number of
particles gives rise to fluctuations that may be important in the
front propagation problem. It has been hypothesized in a variety of
systems~\cite{mean-dla,gene-cutoff,virus1} that the 
leading effect of such fluctuations is to provide an
effective cutoff on the reaction rate at very chemical concentrations.
If this is the case, calculations by Brunet and Derrida~\cite{BD} predict that in
the case of a system which
(in the deterministic limit) exhibits (linear) marginal stability (MS) selection ,
the front velocity obeys the scaling
$v_{MS} - v \sim {1\over \ln^2 {N}}$, where $N$ is the (mean-field) number of
particles per site in the stable state. Direct simulations of the 
underlying Markov processes have, in two cases to date~\cite{BD,KNS}, been consistent
with this predicted form, albeit with some uncertainty regarding the
coefficient. Also, we note in passing that the cutoff idea is the simplest
one which explains the recently discovered fact\cite{nature} that one can have
diffusive instabilities of a front in a chemical reaction system which
do not show up in a reaction-diffusion treatment thereof.

Our purpose here is to introduce a different approach for studying the role of
these fluctuations in modifying the front velocity. There is a
well-established machinery which transforms the master equation for Markov processes for
chemical reaction systems to the solution of an associated stochastic
differential equation. This was first proposed by Doi\cite{Doi}, and clarified in
some seminal work of Peliti~\cite{Peliti}. This framework has in fact been used for
the study of critical phenomena associated with bulk transitions in
reaction dynamics~\cite{CT}, but has not been applied to the issue of front propagation
far from such a bulk transition. Here, we directly simulate the
relevant stochastic equation; this requires the analytic solution of
a (interesting in its own right) single-site problem, which then, via
a split-step method, allows up to time-step the entire spatially-extended
system. Our results to date verify the Brunet-Derrida scaling and in
fact are even consistent with the coefficient obtained by the
cutoff approach.

The outline of this work is as follows. In section II, we review the
mapping from the master equation to a Langevin equation with multiplicative
noise. Next, we solve a variety of single-site problems as a prelude to
introducing our simulation method. We then tackle the front problem
numerically and compare our findings to the results obtained by augmenting
the deterministic system with a cutoff. In order to accomplish this,
the findings of Brunet and Derrida are extended to include the effects
of finite resolution in space and time. At the end, we summarize the open
issues that we hope to address in the future.

\section{Derivation of the stochastic equation}

In this paper, we will study the following space-lattice reaction scheme:
\begin{equation}\label{split}
A \stackrel \alpha \rightarrow 2A \nonumber
\end{equation}
\begin{equation}\label{collis}
2A \stackrel \beta \rightarrow A \nonumber
\end{equation}
\begin{equation}\label{anih}
A \stackrel \lambda \rightarrow 0 \nonumber
\end{equation}
\begin{equation}\label{birth}
0 \stackrel \tau \rightarrow A \nonumber
\end{equation}
\begin{equation}\label{dif}
A_i \stackrel \mu \rightarrow A_e \nonumber
\end{equation}
where $e$ is the nearest neighbor sites of site $i$; $\alpha$, $\beta$, $
\lambda$,
$\mu$, $\tau$ are rates of the corresponding reactions, i.e. probabilities of
transition per unit time. This  process is described by the master 
equation 
\begin{eqnarray} \label{master}
\frac{dP(\{n_i\};t)}{dt}=\sum_{i} \left[ \frac{\partial P(\{n_i\};t)}{\partial t}
\bigg\vert_{\alpha}+
\frac{\partial P(\{n_i\};t)}{\partial t}\bigg\vert_{\beta}+
\frac{\partial P(\{n_i\};t)}{\partial t}\bigg\vert_{\lambda} \right. +\nonumber\\
\left. \frac{\partial P(\{n_i\};t)}{\partial t}\bigg\vert_{\tau}+
\frac{\partial P(\{n_i\};t)}{\partial t}\bigg\vert_{\mu} \right]
\end{eqnarray}
which states that the probability $P(\{n_i\};t)$ of having $n_i$ particles 
on sites
$i$ at some time $t$ changes via each of the elementary processes:
\begin{enumerate}
\item one particle splitting into two 
\begin{equation}\label{massplit}
\frac{\partial P(\{n_i\};t)}{\partial t}\bigg\vert_{\alpha}=\alpha[(n_i-1)
P(\ldots,n_i-1,\ldots;t)-n_i P(\ldots,n_i,\ldots;t)]
\end{equation}
\item
two particle reaction with one being annihilated
\begin{equation}\label{mascol}
\frac{\partial P(\{n_i\};t)}{\partial t}\bigg\vert_{\beta}=\beta[(n_i+1)n_i
P(\ldots,n_i+1,\ldots;t)-n_i(n_i-1)P(\ldots,n_i,\ldots;t)]
\end{equation}
\item
one particle annihilation 
\begin{equation}\label{masanih}
\frac{\partial P(\{n_i\};t)}{\partial t}\bigg\vert_{\lambda}=\lambda[(n_i+1)
P(\ldots,n_i+1,\ldots;t)-n_i P(\ldots,n_i,\ldots;t)]
\end{equation}
\item
particle birth from vacuum 
\begin{equation}\label{masbirth}
\frac{\partial P(\{n_i\};t)}{\partial t}\bigg\vert_{\tau}=\tau[P(\ldots,n_i-1,
\ldots;t)
-P(\ldots,n_i,\ldots;t)]
\end{equation}
\item
diffusion 
\begin{equation}\label{masdif}
\frac{\partial P(\{n_i\};t)}{\partial t}\bigg\vert_{\mu}=\mu\sum_{e}[(n_e+1)
P(\ldots,n_i-1,n_e+1,\ldots;t)-n_i P(\ldots,n_i,\ldots;t)]
\end{equation}
\end{enumerate}
In this section, we provide a self-contained derivation of the
stochastic equation whose solution is directly related to the solution
of this master equation. This is by now fairly standard, but we find
it useful to include this derivation here both for completeness and
for fixing various parameters in the final Langevin system.
 
Following Doi\cite{Doi}, 
we introduce a vector in Fock space $|\{n_i\}\rangle$ and raising and lowering
operators $\aa^+_i$, $\aa_i$ with the properties:
\begin{eqnarray}
\aa_i|\ldots, n_i, \ldots\rangle =n_i |\ldots, n_i-1, \ldots\rangle \nonumber\\
\aa_i^+|\ldots, n_i, \ldots\rangle = |\ldots, n_i+1, \ldots\rangle 
\end{eqnarray}
and the  commutation relation
\begin{equation} \label{com}
[\aa_i,\aa_j^+]=\delta_{ij}
\end{equation}
We choose an initial condition for the master equation $(\ref{master})$ 
to be a Poisson state,
\begin{equation}
P(\{n_i\};t=0)=e^{-N_A(0)} \prod_i \frac{n_{0i}^{n_i}}{n_i!}
\end{equation}
where $N_A(0)=\sum\limits_i n_{0i} $ is the expected total number of particles.
If we define the time dependent vector
\begin{equation}
|\phi(t)\rangle=\sum_{\{n_i\}}P(\{n_i\};t)|\{n_i\}\rangle
\end{equation}
the master equation can be written in the Schr\"{o}dinger form
\begin{equation}\label{Schr}
\frac{\partial}{\partial t}|\phi(t)\rangle=-\hat{H}|\phi(t)\rangle
\end{equation}
where $\hat{H} = \sum_i \hat{H}_i$ and the latter is given by
\begin{equation}\label{H}
 -\mu \sum_e \aa_i^+(\aa_e-\aa_i)+
\alpha [1-\aa_i^+]\aa_i^+a_i-
\beta [1-\aa_i^+]\aa_i^+\aa_i^2-
\lambda [1-\aa_i^+]\aa_i+
\tau [1-\aa_i^+] 
\end{equation}
The formal solution of this equation is
\begin{equation}
|\phi(t)\rangle =e^{-\hat{H}t}|\phi(0)\rangle=e^{-N_A(0)}e^{-\hat{H}t}e^{
\sum_i \aa_i^+ n_{0i}}|0\rangle
\end{equation} 

To be able to calculate average values for observables, we need to introduce 
the projection state 
\begin{equation}
\langle |=\langle 0 | \prod_i e^{\aa_i}
\end{equation} 
The external product of this with any state $|\{n_i\}\rangle$ gives $1$.
Then any normal-ordered polynomial operator satisfies 
\begin{equation}
\langle|Q(\{\aa_i^+\},\{\aa_i\})=\langle|Q(\{1\},\{\aa_i\})
\end{equation}
Using this equation we get for any observable
\begin{equation} \label{avr}
\langle A(t) \rangle = \sum_{\{n_i\}} A(\{n_i\})P(\{n_i\};t)=\langle|
\tilde{A} (\{\aa_i\})|\phi(t)\rangle
\end{equation}
where $\tilde{A} (\{\aa_i\})$ is what we obtain by using
the commutation relation to normal order $A$ and thereafter setting $\aa_i^+$ to
$1$.

In order to write a path integral representation for the time
evolution operator, we introduce a set of coherent states ( see \cite{CO} for 
more strict treatment)
\begin{equation}
\| \{z_i\} \rangle = e^{\sum_i z_i \aa_i^+-z_i} | 0 \rangle
\end{equation}
where $z_i$ is complex eigenvalue of $\aa _i$.
In the case of real positive $\{z_i\}$ this states are Poissonian states with
$\langle {n}_i \rangle=z_i$.
By inserting the completeness relation, 
\begin{equation}
1=(\prod_i \int \frac{d^2 z_i}{\pi}e^{-|z_i|^2+z_i+z_i^*})\|z\rangle \langle
z\|
\end{equation}
where $d^2 z_i=d(Re{}z_i)d(Im{}z_i)$, into expression $(\ref{avr})$ we get 
\begin{eqnarray}
\langle|
\tilde{A}(\{\aa_i\})e^{-\hat{H}t}|\phi(0)\rangle=\langle|
\tilde{A}(\{\aa_i\})(1-\hat{H}\Delta t)^{N \Delta t}\|z^{(0)}\rangle= \nonumber \\
\langle|\tilde{A}(\{\aa_i\})
(\prod_i \int \frac{d^2 
z^{(N)}_i}{\pi} e^{-|z_i^{(N)}|^2+z_i^{(N)}+{z^*_i}^{(N)}})
\|z^{(N)}\rangle \langle z^{(N)}\| \times  \nonumber \\   
(\prod_{j=1}^N (1-\hat{H} \Delta t))
\|z^{(0)}\rangle= 
\langle 0 | (\prod_{j=1}^N \prod_i \int \frac{d^2 z^{(j)}_i}{\pi})
 \tilde{A}(\{z^{(N)}_i\}) e^{-S}|0 \rangle
\end{eqnarray}
where $\Delta t=t/N$, $|\phi(0)\rangle =\|z^{(0)}\rangle$ and
\begin{eqnarray}
S=\sum_{j=1}^{N}\sum_i(H_i({z^*}^{(j)},z^{(j-1)})\Delta t +
|z_i^{(j)}|^2 -{z_i^*}^{(j)} z_i^{(j-1)}-z_i^{(N)}
+z_i^{(0)})= \\
\label{daction}
\sum_{j=0}^{N-1}\sum_i\Delta t(\frac{\bar{z}_i^{(j+1)}(z_i^{(j+1)}-
z_i^{(j)})}
{\Delta t} + H_i(\bar{z}^{(j+1)}+1,z^{(j)}))
\end{eqnarray}
Here ${z_i^*}^{(j)}=\bar{z}_i^{(j)}+1$ and $H_i(\{{z_i^*}^{(j+1)}\},
\{z^{(j)}\})$
is the same function of ${z_i^*}^{(j+1)}$, ${z_i^*}^{(j)}$ as 
$\hat{H}_i(\{\aa_i^+\},\{\aa_i\})$ of $\aa_i^+$, $\aa_i$.
In the continuous time limit we get 
\begin{equation}
\langle A(t_0) \rangle = \frac{\int \prod_i {\cal D} \bar{z}_i {\cal D} 
z_i 
A(\{z_i(t_0)\})e^{-S[\{\bar{z}_i(t)\},\{z_i(t)\};t_0]}}{\int \prod_i 
{\cal D} \bar{z}_i {\cal D} z_i e^{-S[\{\bar{z}_i(t)\},\{z_i(t)\};t_0]}}
\end{equation}
with  
\begin{eqnarray}
S[\{\bar{z}_i(t)\},\{z_i(t)\};t_0]=\sum_i \int_0^{t_0} dt  \ \left(\bar{z}_i(t)[\frac
{d}{d t} - \mu \nabla ^2 ] z_i(t) 
-\alpha (1+ \bar{z}_i(t)) \bar{z}_i(t) z_i(t) +  \right. \nonumber \\
\left. \beta (1+ \bar{z}_i(t)) \bar{z}_i(t) z_i^2(t) +
\lambda  \bar{z}_i(t) z_i(t) -
\tau  \bar{z}_i(t) \right)
\end{eqnarray}
where $\nabla ^2$ is the lattice Laplacian $ z_i(t) =\sum_e (z_e(t) - z_i(t))$.
Now, we linearize the action  using the Stratonovich transformation 
\begin{equation}
e^{\bar{z}_i^2(\alpha z_i -\beta  z_i^2) dt} \sim
\int d \eta_i e^{-\frac{1}{2}\eta_i^2- \bar{z}_i \sqrt{2(\alpha z_i 
-\beta  z_i^2)} \eta_i \sqrt{dt} }
\end{equation}
and integrate out the $\bar{z}$ variables
\begin{eqnarray}
\langle A(t_0) \rangle \sim \int \prod_i {\cal D} \eta_i {\cal D} 
z_i e^{-\frac{1}{2}
\sum_i \int_0^{t_0} dt \eta_i^2(t)}A(\{z_i(t_0)\}) 
\prod_{t=0}^{t_0}\delta ( dz_i(t) - \mu \nabla^2 z_i(t) dt -  \nonumber\\ 
\alpha z_i(t)dt +
\beta  z_i^2(t) dt + \lambda z_i(t) dt - \tau dt - \sqrt{2(\alpha z_i(t) 
-\beta  z_i^2(t))} \eta_i(t) \sqrt{dt}) 
\end{eqnarray}
In this expression, there are  $\delta$-functions at every time; this means that
only $z_i(t)$ which satisfy to the Langevin equation
\begin{equation} \label{Langevin}
dz_i(t)= \mu \nabla^2 z_i(t) dt + 
(\alpha- \lambda) z_i(t)dt -
\beta  z_i^2(t) dt  + \tau dt + \sqrt{2(\alpha z_i(t) 
-\beta  z_i^2(t))} dW_i(t)
\end{equation}
(where $W_i(t)$ is a Wiener process)
contribute to the path integral. In other words, the variables $z_i(t)$ remain
on the trajectories described by equation $(\ref{Langevin})$.
Note that this equation must be considered as an Ito stochastic differential 
equation, since we can see from the form of the action  $(\ref{daction})$
that the updating the variables $z_i$ to time-step $j+1$ only requires knowledge
of the variables at time-step $j$. Also, we note that for $\lambda
\geq 0$ and for small enough (positive) $\tau$, if the initial conditions
specify $0\leq z_i(0)\leq\alpha/\beta$, this will remain true for
all subsequent time. Thus, equation $(\ref{Langevin})$ describes the
temporal evolution of the system as sequence of Poissonian states~\cite{Gardiner}.

For further analysis, we rescale $(\ref{Langevin})$ with $z=u \alpha/\beta$,
$t\rightarrow t/\alpha$, $\tilde{\lambda}=\lambda/\alpha$, $\tilde{\tau}=
\tau \beta /\alpha^2$, and $\tilde{\mu}=\mu/\alpha$. If we
furthermore let $N=\alpha / \beta$ be the mean-field number of particles in the
presence only the first two processes (no spontaneous decay or spontaneous
creation),  we obtain
\begin{equation} \label{L}
du_i=  \tilde{\mu} \nabla ^2 u_i dt + 
(1- \tilde{\lambda})u_idt -
u_i^2 dt  + \tilde{\tau} dt + \sqrt{\frac{2}{N}} \sqrt{u_i 
-u_i^2} dW_i
\end{equation} 
with initial conditions $0 \le u_i (0) \le 1$.

\section{Exact Solutions of Some Local Langevin Equations}

In the absence of process $(\ref{birth})$, i.e. at $\tau=0$, equation $(\ref{L})$
has an absorbing state $u=0$.
In the vicinity of this point,  equation $(\ref{L})$ cannot be treated by
merely setting $dt$ to a finite time-step. Such a scheme would often
give rise to a negative $u$, due to the (very-large) noise term. 
One ad-hoc way to circumvent this difficulty was given by Dickman~\cite{Dickman}, who proposed 
to re-introduce discreteness into the state-space in the vicinity of the
absorbing state. Although this approach appears to work (it seems to lead
to the correct critical behavior near the bulk transition of this class
of models), it seems to be a step backward; after all, the original process
was discrete and the whole purpose of using the Langevin formalism is to
provide a (hopefully more analytically tractable) continuum description.
But, one must then come up with a different scheme for updating the
stochastic variables.

Our approach is to solve exactly the stochastic part of the evolution
equation and embed this via the split-step method in a complete update
scheme for a finite time-step $\Delta t$. We will discuss the details
of this scheme in the next section. Here, we provide an analytic
solution for several (local) Langevin equations, as these results will be
needed later. Also, this solution set is of interest on its own.  There is
some limited consideration of equations of this sort in the 
literature~\cite{SQRT}, but
as far as we can determine, these explicit solutions for the case of physical
no-flux boundary conditions at the absorbing state has not previously appeared.

So, 
we consider Langevin equations with just the noise term. Let us start with
the simplest example, 
\begin{equation} \label{Lsqrt}
du=\sqrt{2 u} dW
\end{equation} 
The probability density $P(u,t)$ satisfies the associated Fokker-Planck equation 
\begin{equation} \label{FP}
\frac{\partial P(u,t)}{\partial t}=\frac{\partial^2}{\partial u^2} u P(u,t)
\end{equation}  
with initial condition $P(u,t) \bigg\vert_{t=t_{in}}=\delta(u-u_0)$. We want
our solution to be equal to zero at $u<0$ and have no flux leaking out
of this point; this will guarantee that the total probability remains a
constant, which we will choose to be unity.

To solve this equation, we define $\psi = uP$ and Laplace transform in time to obtain
\begin{equation}
u  \frac{\partial ^2 \tilde{\psi} } {\partial u^2} (s)  \ - \ 
s \tilde{\psi} (s)  \ = \ 
u_0 \delta (u-u_0)
\end{equation}
Here, $\tilde{\psi} (s)$ is the transform of $\psi$.
If we let $y=2\sqrt{u}$, this can be written as
\begin{equation}
\left( \frac{\partial^2}{\partial y^2} - 
\frac{1}{y}\frac{\partial}{\partial y} - s \right) \tilde{\psi} \ = \ 
-\frac{y_0}{2} \delta (y-y_0)
\end{equation}
The homogeneous part of this equation can be recognized as a variant of
Bessel's equation. This allows us to write down a provisional solution in the form
\begin{equation}
\tilde{\psi} (s) \ =  \ \frac{y y_0}{2} K_1 (\sqrt{s} y_> ) I_1 (\sqrt{s} y_<)
\end{equation}
where $I_1$ and $K_1$ are modified Bessel functions and $y_> (y_<)$ is
the larger (smaller) of $y$ and $y_0$. Returning to the original variables,
\begin{equation}
\tilde{P} (s) \ =  \ 2 \frac{\sqrt{u_0}}{\sqrt{u}} K_1 
(2 \sqrt{s u_> }) I_1 (2\sqrt{s u_<} )
\end{equation}
This solution does not, of course,
vanish for $u<0$ and hence we must modify it by multiplying by $\theta (u)$.
This does not change the fact that it solves the equation away from $u=0$
but it does introduce a discontinuity of size $2 \sqrt{u_0}  K_1 (2 \sqrt{s u_0 })
 $. If we look at the original equation, we see that
this leads to a $\delta$ function via $u \delta ' (u) = -\delta (u)$. This
must be compensated by adding an explicit $\delta$ function piece to the
solution. The final result is
\begin{equation}
\tilde{P} (s) \ =  \ 2 \frac{\sqrt{u_0}}{\sqrt{u}} K_1 
(2 \sqrt{s u_> }) I_1 (2\sqrt{s u_<} )  \\
 + \ 2 \sqrt{\frac{u_0}{s}}  K_1 (2 \sqrt{s u_0 }) \delta (u)
\end{equation}

We can do the inverse transform by the usual contour integral approach.
The details are particularly unilluminating, so we merely quote the final
result
\begin{equation}\label{solsqrt}
P (u,t)  \ =  \  \frac{1}{t}  \sqrt{\frac{u_0}{u}} e^{-(u+u_0) /t} 
I_1 (\frac{2\sqrt{u u_0}}{t} ) 
 + e^{-u_0/t}  \delta (u)
\end{equation}
One can check explicitly that this solves the equation and also that
$P$ remains normalized for all times. The $\delta$ function piece
represents accumulation at the absorbing state; as $t$ gets large, all
the probability end up there. The regular part of
$P(u,t)$ is presented
on the Figure \ref{f1}. We see that as ratio $u_0/ t$ becomes 
smaller,
the distribution gradually shifts towards zero and differs
from the Gaussian expected at very short times.

For completeness, we write down the solution of Langevin equations 
with additional terms. If we take the system,
\begin{equation} \label{tau}
du=\tau dt + \sqrt{2 u} dW ,
\end{equation}  
the probability density is 
\begin{equation}
P(u,t) \ = \ \left( \frac{u_0}{u} \right) ^{\frac{1-\tau}{2}}  \ 
\frac{I_{\tau-1}(2 
\frac{\sqrt {u u_0}}{t}) }{t} \ e^{-(u+u_0)/t}
\end{equation}
For this case, spontaneous birth from the vacuum prevents the system from 
falling irreversibly to the state $u=0$. Instead, there is an integrable
power-law singularity  near $u=0$ which becomes a $\delta$ function in the 
$\tau \rightarrow 0$
limit; this is shown in Figure \ref{f2}.

For the system 
\begin{equation}
du=\alpha u dt + \sqrt{2 u} dW \label{alpha}
\end{equation}  
the probability density is 
\begin{equation}
P(u,t) \ = \ e^{-\frac{\alpha u_0 e^{\alpha t}}{e^{\alpha t }-1}}
\delta (u) \ + \ \sqrt{g(t) \frac{u_0}{u}} 
I_1 \left( 2 
\sqrt {g(t) u u_0}\right)  \ e^{-\alpha \frac{u+u_0 
e^{\alpha t }} {e^{\alpha t }-1}}
\end{equation}
where $g(t)=\frac{\alpha^2 e^{\alpha t}}{ (e^{\alpha t }-1)^2}$. In this
case, the drift toward infinity gives rise to a finite total probability
(for long times) of falling into the absorbing state. One can also work
out the case of both finite $\tau$ and finite $\alpha$.

For the system 
\begin{equation}\label{u-u2}
du=\sqrt{2 (u-u^2)} dW
\end{equation}
we can derive a series representation for the  
probability density, 
\begin{eqnarray}\label{full}
P(u,t)=
\sum_{m=1}^{+\infty} e^{-m(m+1)t} \frac{2m+1}{2m(m+1)} \sqrt{\frac{
u_0(1-u_0)}{u(1-u)}}P_m^1(2 u_0 -1)P_m^1(2 u -1) + \nonumber\\
\delta(u)[1-u_0-\sum_{m=1}^{+\infty}(-1)^m\frac{2m+1}{2m(m+1)}\sqrt{u_0(1-u_0)}
e^{-m(m+1)t} P_m^1(2 u_0 -1)] - \nonumber\\
\delta(1-u)[u_0 + \sum_{m=1}^{+\infty}\frac{2m+1}{2m(m+1)}\sqrt{u_0(1-u_0)}
e^{-m(m+1)t} P_m^1(2 u_0 -1)]
\end{eqnarray}
where  $P_m^1(x)$ is Legendre polynomial. Successive terms in this sum decay
rapidly because of the fast exponent, and the sum can be computed numerically
to high accuracy. Note that there are absorbing states at both $u=0$ and
$u=1$. If the initial state starts close to one of these, the probability
is almost the same as $(\ref{solsqrt})$; if the initial state is in between,
then for short times the system is almost Gaussian.

\section{Front propagation}

\subsection{Numerical method}

We are interested in numerically solving equation $(\ref{Langevin})$,
for the particular case of $\tau = \lambda =0$; this is case which
reduces in the deterministic case to the well-studied Fisher equation{\cite{genetics}  For
this purpose,  we define a function 
\begin{equation} 
F_{u_0,\Delta t}(u)=\int_{0^-}^u P(u,\Delta t)du
\end{equation}
where $P(u,\Delta t)$ is the analytical solution of a single-site Langevin 
equation such as 
$(\ref{L})$.  This function has values ranging from $0$ to $1$. If $y$ is random
variable homogeneously distributed on $[0,1]$, then 
$u=F^{-1}_{u_0,\Delta t}(y)$
is distributed according to corresponding truncated Langevin equation at
time $\Delta t$. The remaining parts of
complete Langevin equation are deterministic and for those terms we can update $u$ via
a simple Euler scheme. We then can combine this two steps together; we first 
compute the 
change in $u$ due to fluctuations and then the change of $u$ due to the
deterministic part (using new value of $u$). Thus 
\begin{equation}
u(t+\Delta t)=F^{-1}_{u(t),\Delta t}(y) + {\cal D}\{F^{-1}_{u(t),\Delta t}(y)\}
\Delta t
\end{equation}
where  ${\cal D}\{u \}$ denotes the terms remaining after consideration of the
noise term.

It is important to note that this scheme never allows
the field to go below zero, but it does allow a variable to be
stuck at zero until it is ``lifted" by the diffusive interaction. This
is an absolutely necessary aspect of simulating processes with an absorbing
state. Approximations which do not allow for getting ``stuck", such as
the system-size expansion method of Van-Kampen~\cite{Kampen} (where the noise correlation
is taken to be related to the solution of the deterministic limit of the
equation) get this wrong and hence cannot get the correct front velocity.
This explains why the simulation results of \cite{LLM} do not at all exhibit
the anomalous $N$ dependence expected via the Brunet-Derrida cutoff argument.
As we will see, our approach is much more successful.

\subsection{Marginal stability criterion for a discretized Fisher equation}
\label{A1}

As $N$ gets large, our results should approach those of the deterministic
system. Since this problem corresponds to propagation into an unstable
region, the velocity should be given by the marginal stability approach.
As is well known, this predicts a velocity equal to $2 \sqrt{D}$,
in the continuum (in time and space) limit.
Here, we extend this result to a discrete lattice and a finite time
update scheme, so as to be able to directly compare our simulation data
with the theoretical expectation.

After linearization, the  deterministic part of the discretized 
equation takes the form
\begin{equation} \label{fd}
\frac{u_i^{(j+1)}-u_i^{(j)}}{\Delta t}= \mu (u_{i+1}^{(j)}-2 u_i^{(j)} +
u_{i-1}^{(j)}) + u_i^{(j)}
\end{equation}
We want to compare this equation to the usual Fisher equation with diffusion coefficient
$D$. This means that $\mu=D/h^2$; we will consider the case $D=1$. We assume
that the front moves with constant velocity $c$ and therefore the variables 
$u_i^{(j)}$ show a stroboscopic picture of this motion at times $j \Delta t $ on the lattice
sites $i$. If we move with the speed of the front  we will see that its shape
exponentially decays as $e^{q (ih-c j \Delta t)}$. Substitution of this 
expression to $(\ref{fd})$ gives the dependence of $c$ on the decay rate $q$
\begin{equation}\label{disp}
\frac{e^{-q c \Delta t}-1}{\Delta t}\ = \ \frac{2}{h^2}(\cosh q h -1) +1 
\end{equation}
The standard  marginal stability argument predicts that we can determine
the decay rate and  asymptotic speed of the front (for 
a sufficiently localized initial state) by solving 
$(\ref{disp})$ as well its derivative with respect to $q$
\begin{equation}\label{ms}
c e^{-q c \Delta t}\ = \ - \frac{2}{h}\sinh q h
\end{equation}
Simulations directly confirm this formula as well as
the  Brunet and Derrida~\cite{BD} result 
(actually derived earlier by Bramson~\cite{Bramson}) that
$c_\infty-c(t) \sim 1/t$ (see Figures \ref{fas01}).

As already mentioned, it has been conjectured that the leading effect
of the fluctuations is the imposition of an effective cutoff of order $1/N$ in the
deterministic equation. To check this, we need to extend the
Brunet and Derrida result to our discretized equation. The basic idea
is that there must be a small imaginary part of the decay rate so
at to satisfy the continuity conditions at the cutoff point; this
is discussed in detail in \cite{BD}. This leads directly to
\begin{equation}
Im \ q\ = \ \frac{\pi q_0}{\ln (A N) }
\end{equation}     
where $q_0$ is solution from the marginal stability criterion and $A$ is some
constant. Since $c'(q)$ is zero at the marginal stability point, we
can find the change in velocity  $\delta c$  by considering the second 
derivative of the function $c(q)$ given by $(\ref{disp})$.
We thus get
\begin{equation} \label{delc}               
\delta c = \frac{\pi^2 q_0 (e^{-c_0 q_0 \Delta t} \cosh q_0 h - 
\frac{c_0^2 \Delta t}{2})}{\ln^2 N} 
\end{equation} 
Again, simulations confirm this formula (see Figure \ref{fcut}). 

\subsection{Results} 

We now present the results of our simulation. We chose to make one
further simplification. We use the pure square root noise term instead
of the precisely correct term given in equation (\ref{L}). We do
this for computational ease, inasmuch as the expression derived for
this case is much simpler than that of equation (\ref{full}). Since it
is only the effect of the noise near the $u=0$ absorbing state which is
crucial for altering the selected velocity, this simplification
should not be essential. Once we have done this, the resulting
equation has the  nice feature that the coefficient $1/\sqrt{N}$ 
in front of the noise term can be removed by the re-scaling  $\hat{u}=u N$.
This means that we can 
simulate equation (\ref{L}) using a fixed probability table (with the same time
step) for any $N$. 

To actually evaluate the probability table $F_{u_0,\Delta t}(u)$, we chose
$512$ equidistant values of $\hat{u}_0$  in the 
interval from 0 to 30. For each $\hat{u}_0$,  the interval of 
values for $\hat{u}$ where  $F_{\hat{u}_0,\Delta t=0.01}(\hat{u})$ 
is non-trivial was divided into $1024$ equidistant points.
The new value of 
$\hat{u}$
was then determined by linear interpolation of the data from the table. For 
$\hat{u}_0 > 30$, new values of  $\hat{u}$ were determined using a standard 
algorithm for the Gaussian distribution , since this distribution is 
the asymptotic limit of equation $(\ref{solsqrt})$, when  $\hat{u}_0 \gg 
\Delta t$, 
$\hat{u}\gg \Delta t$.
The difference for this distribution and exact solution is small 
for $\hat{u}_0>30$.   
Finally, the  computation of the stochastic term was turned off for 
$\hat{u}_0>10^{-3} N$. This should not affect the speed which, we have
already argued, is only sensitive to what happens near $u=0$; 
this insensitivity was also checked directly by running some simulations in
which the stochastic term is included for all values of $u$.

All of our simulations were run up to the time $t=10^{6}$. Four
values of the velocity, corresponding to time intervals of 
approximately $2. x 10^{5}$ were obtained so as to get an average and
some error bar. In Figure \ref{h}. we show data in the form
of $c-c_{\infty}$, where $c_{\infty}$ is calculated from equation (\ref{ms}),
versus $\ln^2{N}$. Also plotted for comparison is the function $ 11/\ln^2{N}$.
Note that over many orders of magnitude of $N$, the dependence derived
by Brunet and Derrida provide a very good fit to the data. 

To get a more accurate indication of the data for large $N$, we present
in Figure \ref{fh1} a series of three runs, for differing values of the spatial lattice
spacing $h$. Under the hypothesis that the stochastic system should be
precisely the same as the deterministic system with the cutoff added,
the expected limiting values at infinite $N$ are shown as triangles
on the axis. It is clearly impossible to definitively conclude that the
curves are approaching these values. On the other hand, simple 
extrapolations come very close and we believe that it is more likely than
not that this hypothesis is true. This is opposite to what was conjectured
based on simulations of a discrete Markov process, where the velocity
seems to scale albeit with a different coefficient. Given the incredibly
slow convergence of this velocity at large $N$, we are pessimistic as to
whether any purely simulational strategy would provide a definitive answer
to this question. This therefore
offers a crucial issue for future theoretical analysis to investigate.

\section{Summary}

In this work, we have shown how to use the field-theoretic mapping of
discrete Markov processes to stochastic equations for continuous density
variable to address the role of finite-particle number fluctuations on
the velocity of reaction fronts. Specifically, we studied a model which
leads to the well-known Fisher equation in the $N \rightarrow \infty$ limit,
where $N$ is the average number of particles per site in the stable state.
Our goal is to understand how the usual marginal stability criterion
becomes modified by these stochastic effects.

It is clear that having finite $N$ lowers the velocity at which a front (corresponding to the 
invasion of the unstable state by the stable one) will propagate. One
attractive hypothesis is that the leading effect of the fluctuations is
to to introduce an effect cutoff into the deterministic equation; this
idea arose independently in model of biological 
evolution~\cite{gene-cutoff} and in mean
field approaches for diffusion-limited-aggregation (DLA)~\cite{mean-dla}. Brunet and
Derrida have shown that if this is the case, one should expect
$v_{MS} - v = {C \over \ln^2 {N}}$, where $C=\pi ^2$ for the case of
continuous time and space. We have extended the calculation of
$C$ to the finite lattice size, finite time-step system and compared this
prediction with direct simulations of the relevant stochastic equation.
Our results verify the form of the scaling and suggest that the
coefficient may be correct as well.

One issue that is left unaddressed by our work to date concerns the
effects of higher spatial dimensionality. It is likely,
although unproven, that the velocity change will be smaller, as
the fluctuations get averaged over the transverse directions. This
seems to be the explanation for the findings of Riordan et al~\cite{doering} that
the reaction front looks mean-field like even for small $N$, in
three and four dimensions. We hope to report on this issue in the future.

Finally, we point out yet again that there is no analytic treatment
available for the velocity selection problem in the stochastic equation.
Obvious expansion methods such as the system-size approach cannot work,
as they neglect the essential role of the fluctuations to push the
system back into the absorbing state at small density. We need to find
a more powerful approach!

We thank D. Kessler for many useful discussions. We also acknowledge the support of the 
US NSF under grant DMR98-5735.

\begin{figure}\centerline{\epsfxsize = 6.in \epsffile{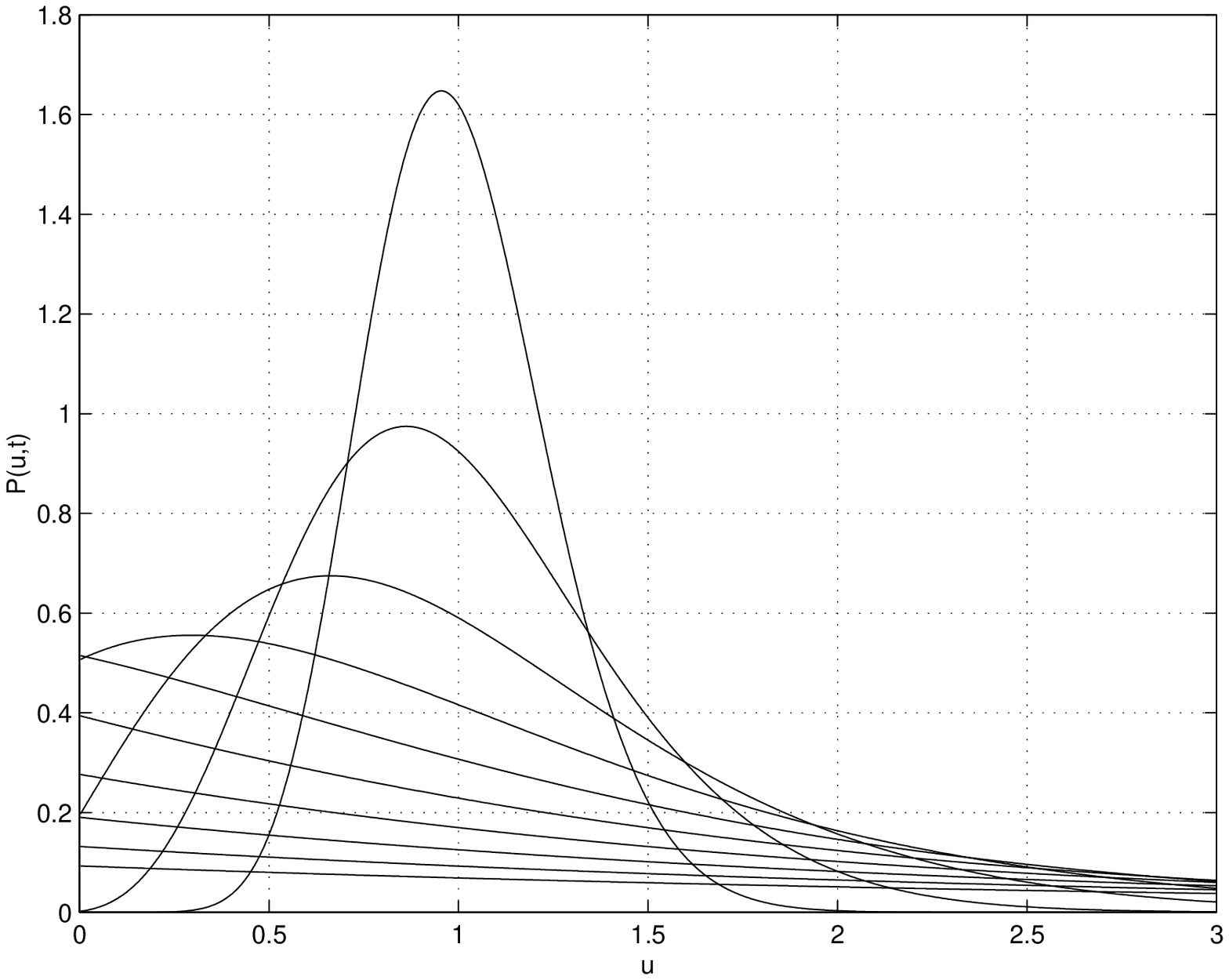}}
\caption{Function $(\protect\ref{Lsqrt})$ with $u_0=1$ and $\Delta t= 0.03(n^2-n+1)$
for $n=1,...,10$.}
\label{f1}
\end{figure}

\begin{figure}\centerline{\epsfxsize = 6.in \epsffile{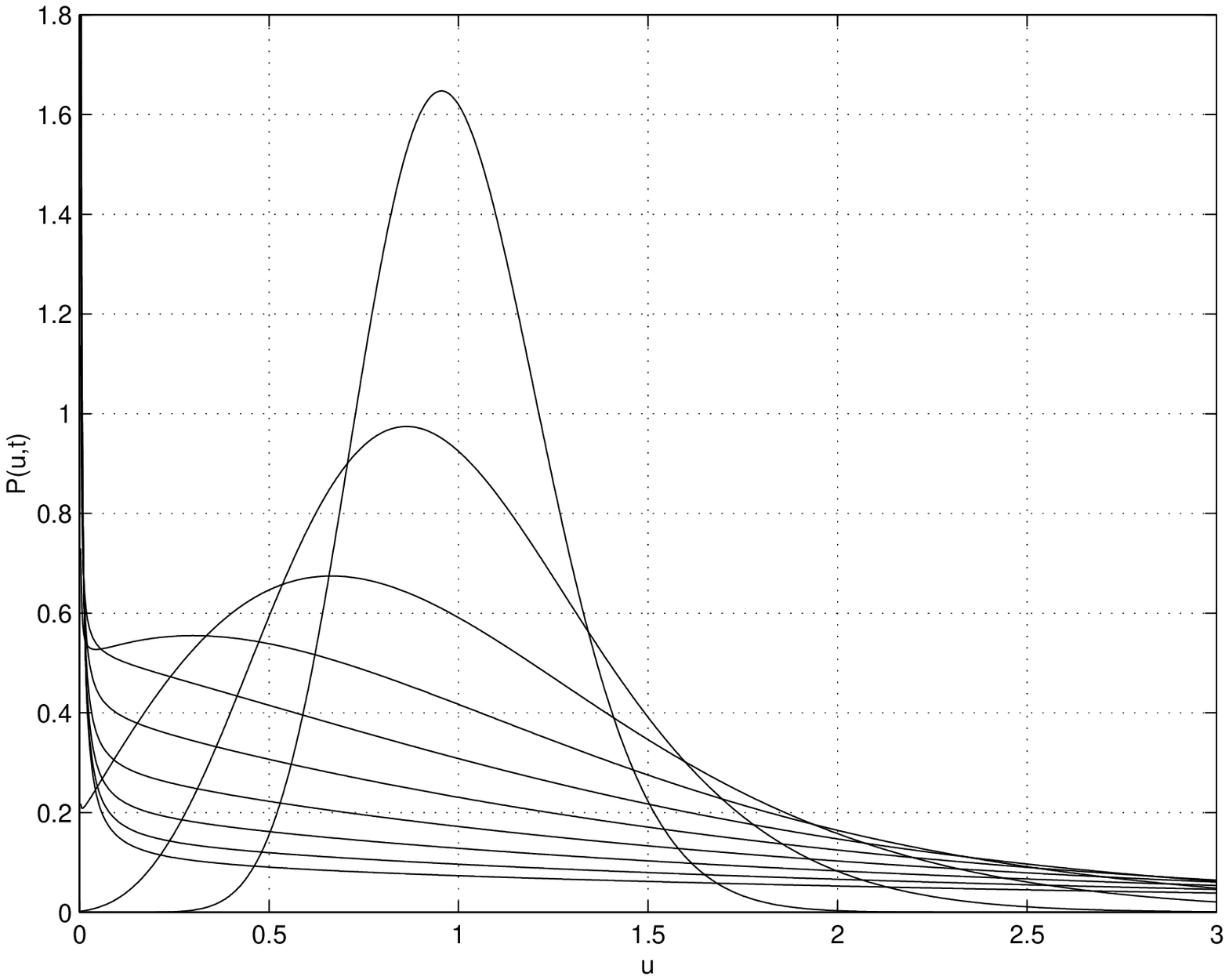}}
\caption{Function $(\protect\ref{tau})$ with $u_0=1$ and $\Delta t= 0.03(n^2-n+1)$
for $n=1,...,10$.}
\label{f2}
\end{figure}

\begin{figure}\centerline{\epsfxsize = 6.in \epsffile{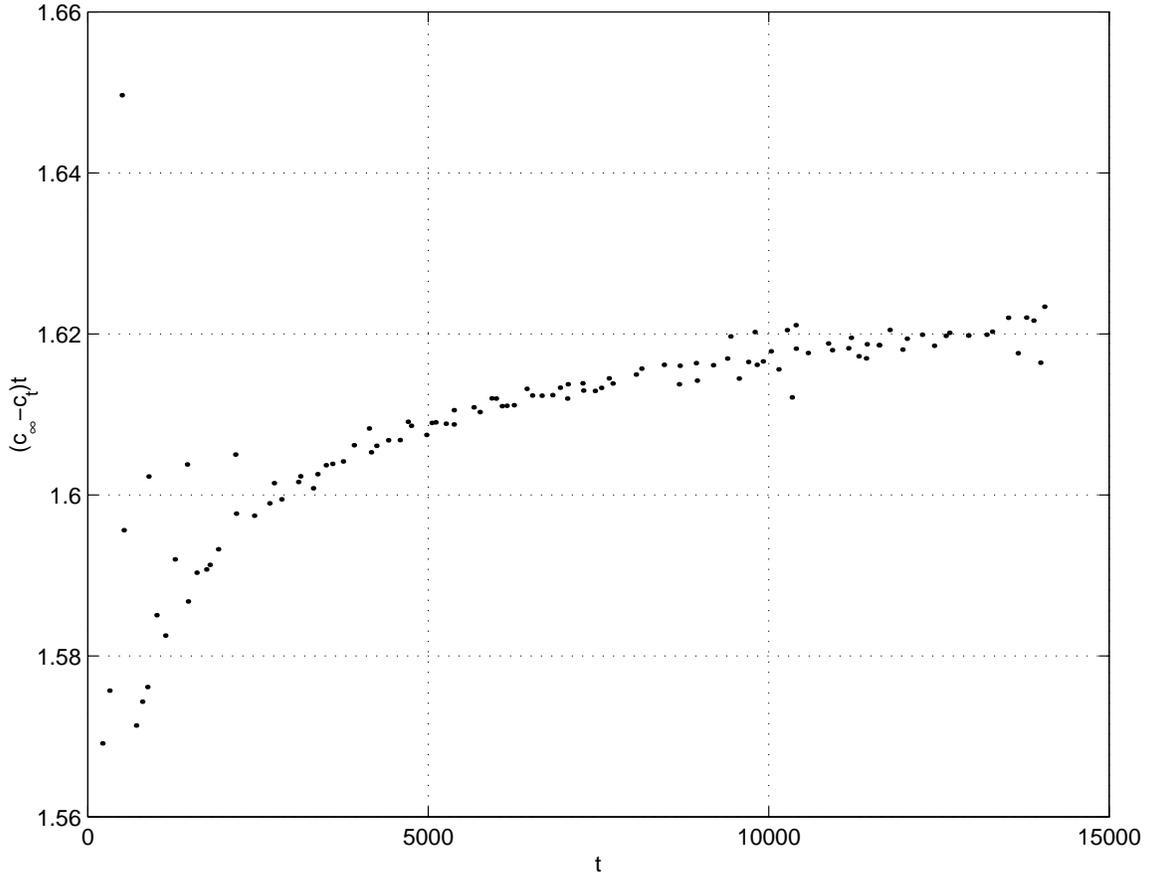}}
\caption{Front propagation for the discretized Fisher equation. Parameters were
$\Delta t=0.01$, $h=1$; for this choice, equation $(\protect\ref{fd})$ gives 
$c_\infty=2.054115884$. The speed is determined by registering every
time some lattice point comes within some small interval around .5
and then merely using the ratio of the number per spaces moved over
the time elapsed; this method leads to some scatter, as seen in
the graph. Aside from showing
agreement with the calculated $c_\infty$, the graph also shows how  
the system approaches the eventual $1/t$ in asymptotic behavior.}
\label{fas01}
\end{figure}

\begin{figure}\centerline{\epsfxsize = 6.in \epsffile{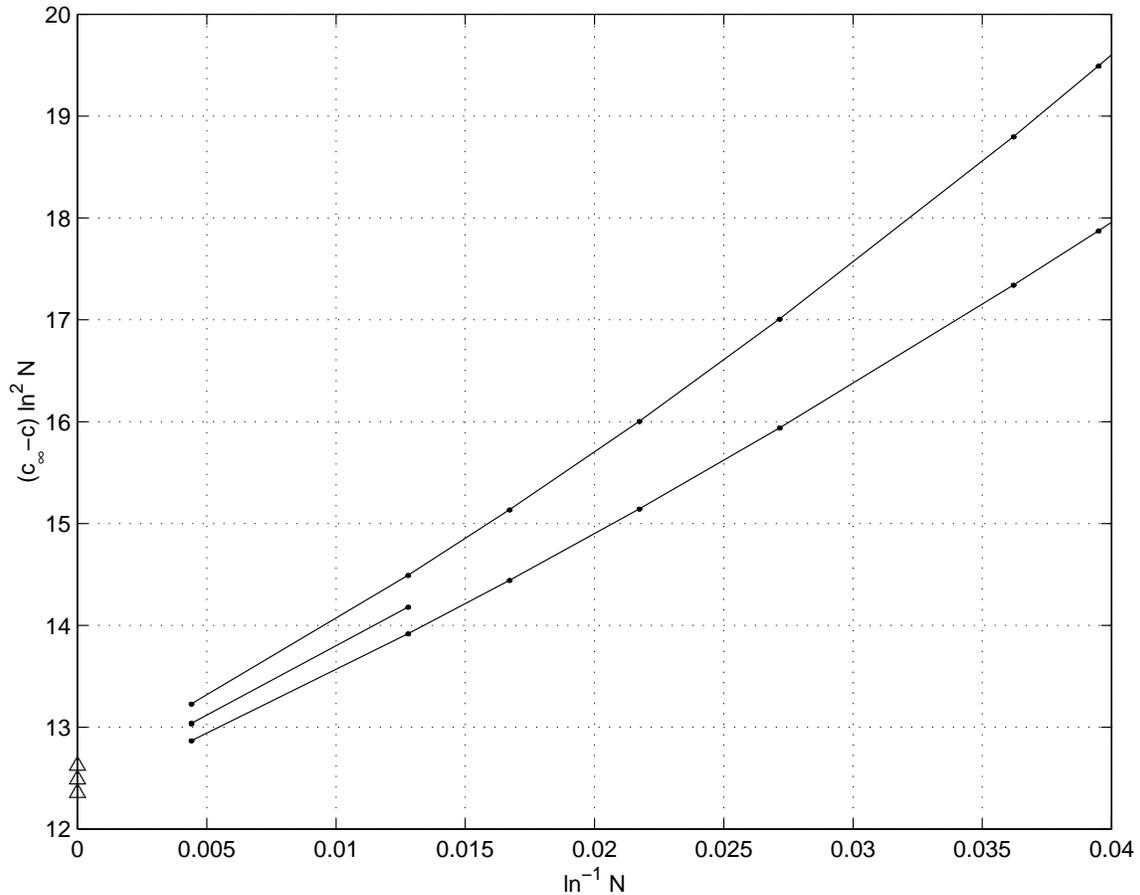}}
\caption{Dependence of the front  speed on the deterministic cutoff. The curves
represent simulation results calculated with the parameters $h=1$ and $\Delta t=0.01$; $0.015$; $0.02$ from top
to bottom respectively. The cutoff was implemented via setting the
field to zero at any time and place where it was below $1/N$. Points on $y$-axis represent theoretical values when
$N \rightarrow \infty$ as  given by equation $(\protect\ref{delc})$.}
\label{fcut}
\end{figure}

\begin{figure}\centerline{\epsfxsize = 6.in \epsffile{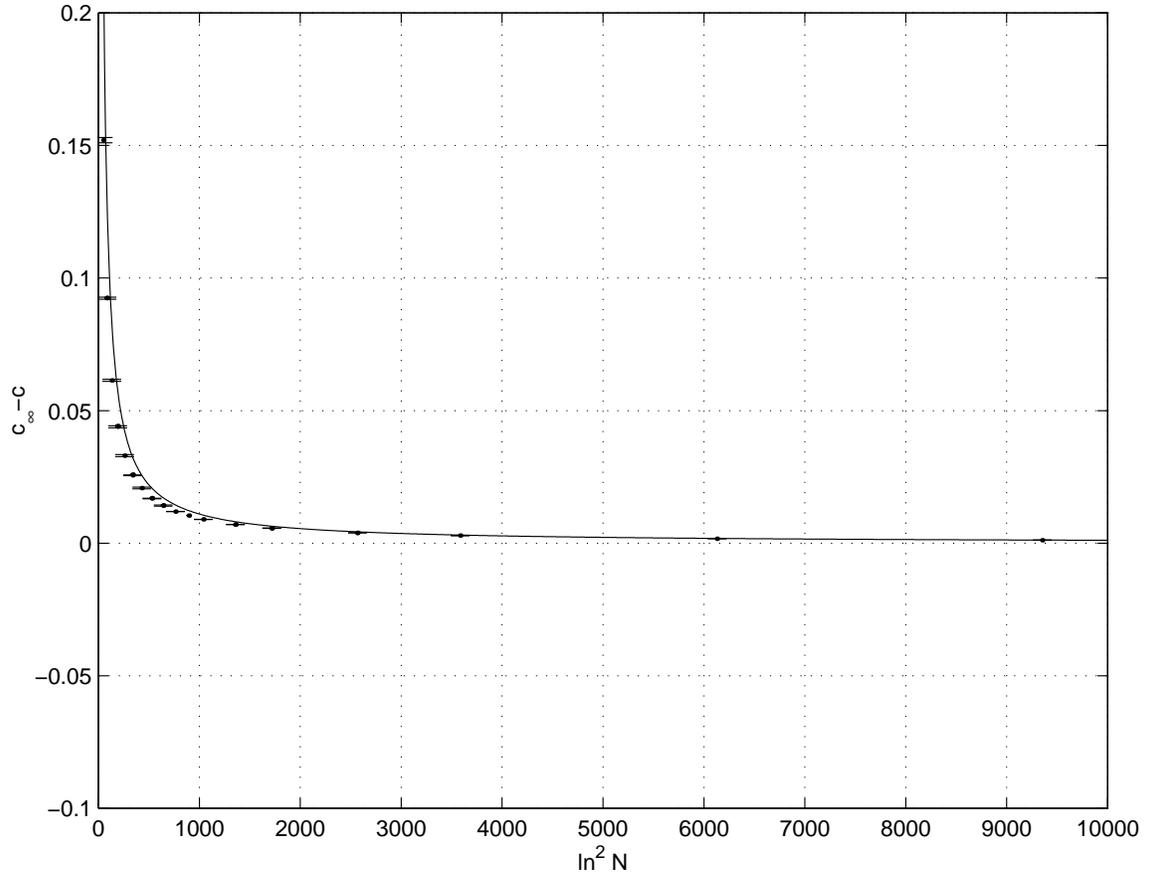}}
\caption{Front speed for the Langevin equation 
with parameters $\Delta t=0.01$ and $h=1$. }
\label{h}
\end{figure}

\begin{figure}\centerline{\epsfxsize = 6.in \epsffile{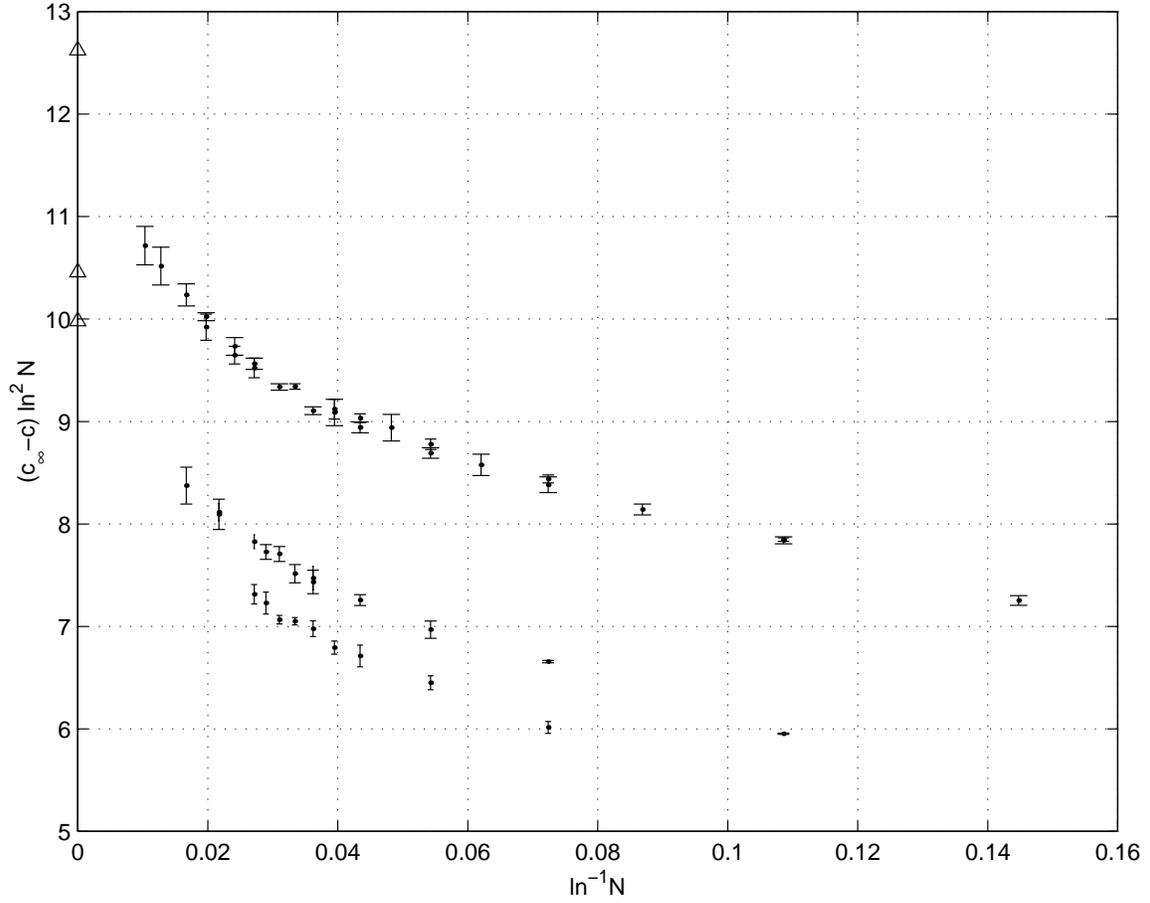}}
\caption{Front speed for the Langevin equation
with parameters $\Delta t=0.01$ and $h=1$, $h=1/2$, $h=1/3$ correspondingly
from top to bottom. Triangles on the $y$-axis are the corresponding 
asymptotic values for the deterministic cutoff.}
\label{fh1}
\end{figure}

\end{document}